\documentclass[twocolumn,showpacs,superscriptaddress,amsmath,amssymb]{revtex4}
\usepackage{amsmath}
\usepackage{dcolumn}
\begin{document}
\title{Realistic shell-model calculations for proton particle-neutron hole nuclei around $^{132}$Sn} % \mbox{\boldmath N=82$}}
\date{\today}
\author{L. Coraggio}
\author{A. Covello}
\author{A. Gargano}
\author{N. Itaco} 
\affiliation{Dipartimento di Scienze Fisiche, Universit\`a di Napoli
Federico II, and Istituto Nazionale di Fisica Nucleare, 
 Complesso Universitario di Monte S. Angelo, Via Cintia, I-80126 Napoli, Italy}
\author{T.T.S. Kuo}
\affiliation{Department of Physics, SUNY, Stony Brook, New York 11794}

\begin{abstract}
We have performed shell-model calculations for nuclei with proton particles and neutron holes around $^{132}$Sn using a realistic effective interaction derived from the CD-Bonn nucleon-nucleon potential. For the proton-neutron channel this is explicitly done in the particle-hole formalism. The calculated results are compared with the available experimental data, particular attention being focused on  the proton particle-neutron hole multiplets. A very good agreement is obtained for all the four nuclei considered, $^{132}$Sb, $^{130}$Sb, $^{133}$Te and $^{131}$Sb. 
We predict many low-energy states which have no experimental counterpart. This may stimulate, and be helpful to, future experiments.

\end{abstract}
\pacs{21.60.Cs, 21.30.Fe, 27.60.+j}

\maketitle

\section{Introduction}

The study of nuclei around doubly magic $^{132}$Sn is a subject of special interest. This is well evidenced by the attention focused on these nuclei in several recent papers \cite{mach01,mach02}, which has resulted in a substantial increase in the knowledge of their properties.  This offers the opportunity for testing the basic ingredients of shell-model calculations, in particular
the neutron-proton effective interaction, well away from the valley of stability. 

During the last few years, we have studied several nuclei in the $^{132}$Sn region in terms of shell model employing realistic effective interactions derived from modern nucleon-nucleon ($NN$) potentials \cite{andr97,cove97,cove99,andr99,cora02}. While our main aim was to assess the ability of such interactions to give an accurate description of nuclear structure
properties, in some cases we have also found it interesting to make predictions which could stimulate experimental efforts to gain further information on nuclei in this region.

So far, we have been mainly concerned with nuclei having few identical valence particles or holes. It is the purpose of this paper to present the results of realistic shell-model calculations for nuclei with proton particles and neutron holes around $^{132}$Sn. 
Actually, the proton particle-neutron hole nucleus $^{132}$Sb was already studied in a previous work \cite{andr99}, where we considered $^{100}$Sn as a closed core and treated the odd proton and the remaining 31 neutrons as valence particles. In that case we derived from the Bonn-A $NN$ potential an effective interaction for two nucleons outside $^{100}$Sn and consistently   
used a unique set of single-particle energies for neutrons and protons (see Ref. \cite{andr99}). This effective interaction, however, may not be quite adequate when moving away from closed shells since many-body correlations are likely to come into play. A more appropriate study of $^{132}$Sb may be performed by making use of the particle-hole ($ph$) formalism, which
implies that the proton-neutron effective interaction has to be explicitly derived for a system with one proton particle and one neutron hole away from  $^{132}$Sn. This is done here starting from the CD-Bonn $NN$ potential \cite{machleidt01} (in the discussion of Sect. IV we will comment on the results of our previous calculations as compared to the present ones).  To have a more complete test of our realistic $ph$ interaction, we also consider the neighboring odd-odd isotope $^{130}$Sb as well as the two 
odd neighbors $^{133}$Te and $^{131}$Sb. We shall see in Sec. III that for the description of the two latter nuclei the effective interaction in the proton-proton and neutron-neutron channels is also quite relevant. The interaction in these two channels is calculated in the particle-particle ($pp$) and hole-hole ($hh$) formalism, respectively.

To place this work in its proper perspective, it should be mentioned that the first attempts to derive realistic effective interactions in the $ph$ formalism
date back to the late 1960s and early 1970s. In this context, we may mention the work of Refs. \cite{kuo68,blom69,kuo70}, where the nuclei $^{208}$Bi, $^{40}$Ca, $^{48}$Ca, and $^{208}$Pb  were studied using $ph$ matrix elements deduced from the Hamada-Johnston potential \cite{Hamada62}.  In Ref. \cite{west69} the   
effective interaction theory was applied to the study of the relation between the particle-particle and $ph$ spectra. An important result of this work was the explanation of the violation of the Pandya $pp$-$ph$ relation \cite{pand56}. Despite these early achievements, little work \cite{skour86} along these lines has been done ever since.

By considering the amount of experimental data which are becoming available on proton particle-neutron hole nuclei around $^{132}$Sn, we have found it timely to revive this kind of calculations making use of a modern $NN$ potential and improved many-body methods for deriving the effective interaction. As regards the latter, we make use here of a new approach \cite{bogner02} which provides an advantageous alternative to the use of the traditional Brueckner $G$ matrix. It consists in deriving a low-momentum $NN$ potential, $V_{low-k}$, that preserves the physics of the original potential $V_{NN}$ up to a certain cut-off momentum $\Lambda$. This is achieved by integrating out  high-momentum components of $V_{NN}$. The scattering phase shifts and deuteron binding energy calculated from $V_{NN}$ are reproduced by $V_{low-k}$. The latter is a smooth potential that can be used directly as input for the calculation of shell-model effective interactions. A detailed description of our derivation of $V_{low-k}$ can be found in Ref. \cite{bogner02}.

Once the $V_{low-k}$ is obtained, the calculation of the matrix elements of the $pp$ and $hh$ effective interaction proceeds in the usual way, as described, for instance, in Refs. \cite{Jiang92,cor00}. Our derivation of the effective interaction in the $ph$ formalism is outlined in Sec. II. In Sec. III we present the results of our calculations and compare them with the experimental data. Section IV contains a discussion and a summary of our conclusions.

\section{Particle-hole effective interaction}

Here, we give a brief description of how to derive the shell-model effective
interaction $V_{\rm eff}$  within the framework of the $ph$ formalism.
We use a model space folded-diagram formalism \cite{kuosnes90}.
The basis states of the model space are the  one-particle one-hole
states \hbox{$\mid \phi_{ph} \rangle \equiv a^{\dagger}_{p} a_{h} 
\mid C \rangle $},
where $a^{\dagger}_{p}$
and  $a_{h}$ denote respectively a proton particle and neutron hole 
creation
operator and $\mid C \rangle$ represents the
$^{132}$Sn core. The secular matrix  is of the form
\begin{equation}
(\epsilon_{p} -\epsilon_{h})\delta_{ph,p'h'}
+ \langle \phi_{ph} \mid  V_{\rm eff} \mid \phi_{p'h'} \rangle,
\end{equation}
where $\epsilon$ denotes the unperturbed single-particle energy.

The effective interaction can be schematically written \cite{kuosnes90} in
operator form as
\begin{widetext}
\begin{equation}
V_{\rm eff} = \hat{Q} - \hat{Q'} \int \hat{Q} + \hat{Q'} \int \hat{Q} \int
\hat{Q} - \hat{Q'} \int \hat{Q} \int \hat{Q} \int \hat{Q} + ~...~~,
\end{equation}
\end{widetext}
where $\hat{Q}$ is the irreducible vertex function, usually referred to
as the $\hat{Q}$-box.
$\hat{Q'}$ is obtained from $\hat{Q}$ by removing terms of first order in
the interaction.  The integral sign above represents a
generalized folding operation.
Once the $\hat{Q}$-box is calculated, $V_{\rm eff}$ is
obtained by summing up the folded-diagram series of Eq. (2) to all orders
by means of the Lee-Suzuki iteration method \cite{suzuki80}.

As mentioned earlier, we have derived from the CD-Bonn $NN$ potential the
$V_{low-k}$ (hereafter this will be abbreviated as $V$).
Since it is a smooth potential (without strong repulsive core), we can
use it directly in the calculation of the 
the vertex function $\hat{Q}$-box, which  is composed of irreducible
valence-linked diagrams. We have included
all such diagrams through second order in $V$, namely
six one-body diagrams and six two-body diagrams, shown respectively
in Fig. 1 and Fig. 2. 

Regarding the one-body diagrams, their calculation within the $ph$ formalism
is the same as that in the $pp$ and $hh$ cases.
However, since we are dealing with both external particle and hole
lines, the calculation of the two-body $\hat{Q}$-box diagrams is
somewhat different.
For example, the familiar core-polarization diagram, see (e) in Fig. 2, is given by

\begin{widetext}
\[
\langle 1~2;J| V_{\rm 2p2h} |3~4;J \rangle  = 
-\frac{1} {\hat{J}} \sum_{J'} \hat{J'} \sum_{\rm ph}
(-1)^{j_2+j_4+j_p+j_h}
X \left( \begin{array}{ccc}
%\vspace{0.1in}
j_3~ j_4~  J  \\ %\vspace{-0.1in}
j_1~ j_2~  J  \\ J'~  J'~  0  \end{array} \right)
\]

\begin{equation}
\times \left[ \frac{\langle 1~p| V | 3~h\rangle_{J'}^{cc}~ 
\langle h~4| V | p~2 \rangle_{J'}^{cc}}{\omega
-(\epsilon_1-\epsilon_4+\epsilon_p-\epsilon_h)}
+ \frac{\langle 1~h| V | 3~p \rangle_{J'}^{cc}~ 
\langle p~4| V | h~2\rangle_{J'}^{cc}}{\omega
-(\epsilon_3-\epsilon_2+\epsilon_p-\epsilon_h)} \right],
\end{equation}
\vspace{0.1in}
\end{widetext}

\noindent
where $\hat{x}=(2x+1)^{1/2}$ and $\omega$ is the so-called starting energy.
$X$ is the standard normalized 9-$j$ symbol.
The matrix elements on the right-hand side of Eq. (3) are 
the cross-coupled ones \cite{kuo81}. They are related to the usual 
direct-coupled 
matrix elements  by the simple transformation
\begin{widetext}
\begin{eqnarray}
\langle 1~2 | V |3~4 \rangle_{J'}^{cc} & = &
\frac{1} {\hat{J'}} \sum_{J''} \hat{J''}~
X \left( \begin{array}{ccc}
j_3~ j_4~  J''  \\ 
j_1~ j_2~  J''  \\ J'~  J'~  0  \end{array} \right) 
\langle 1~2;J''| V | 3~4;J''\rangle.
\end{eqnarray}
\end{widetext}

\noindent 
Using the diagram rules described in Ref. \cite{kuo81},
the expressions for the diagrams in Fig. 1 and the other
second-order  diagrams of Fig. 2 can be readily obtained.

Note that in our calculation
the effective interaction
has both one-body and two-body connected terms. The one-body terms summed to
the unperturbed energies of Eq. (1)
represent the single-particle (SP) and single-hole (SH) energies of $^{133}$Sb
and $^{131}$Sn, respectively.
Thus, in Eq. (1) we have replaced the unperturbed energies with the 
experimental SP and SH energies, while in the
$\langle \phi_{ph} \mid  V_{\rm eff} \mid \phi_{p'h'} \rangle$
matrix elements we have
retained  only the
two-body connected parts for the effective interaction.
A subtraction method \cite{shurpin83} has been used
in extracting the two-body connected parts from the folded diagram series
of Eq. (2).

\section{Results and Comparison with experiment}

In this Section we present and compare with experiment the results of a shell-model study of
the four nuclei $^{132}$Sb, $^{130}$Sb, $^{133}$Te, and $^{131}$Sb. In our calculations we consider $^{132}$Sn
as a closed core and let the valence
protons and neutron holes occupy the five levels $0g_{7/2}$, $1d_{5/2}$, $1d_{3/2}$, $2s_{1/2}$,
and $0h_{11/2}$ of the 50-82 shell.

As mentioned in the Introduction, we start from the CD-Bonn free $NN$ potential and
derive the $V_{low-k}$ using a value of the cutoff parameter $\Lambda$=2.1 fm$^{-1}$
(see Ref. \cite{bogner02}). 
The effective interaction in the proton-neutron, proton-proton, and neutron-neutron channels has been derived in the $ph$, $pp$ and $hh$ representation, respectively.

The single-particle (SP) and single-hole (SH) energies have been taken from the
experimental spectra of $^{133}$Sb \cite{sanc99} and $^{131}$Sn \cite{foge84a,foge84b}, respectively. The only exception is 
the proton $\epsilon_{s_{1/2}}$ which was taken from Ref. \cite{andr97}, since the corresponding SP level has not been observed in $^{133}$Sb.
Our adopted values for the proton SP energies are (in MeV):
$\epsilon_{g_{7/2}}=0.0$, $\epsilon_{d_{5/2}}=0.962$, $\epsilon_{d_{3/2}}=2.439$, 
$\epsilon_{h_{11/2}}=2.793$, and   $\epsilon_{s_{1/2}}=2.800$, and for the neutron SH energies: 
$\epsilon^{-1}_{d_{3/2}}=0.0$, $\epsilon^{-1}_{h_{11/2}}=0.100$, $\epsilon^{-1}_{s_{1/2}}=0.332$, 
$\epsilon^{-1}_{d_{5/2}}=1.655$, and   $\epsilon^{-1}_{g_{7/2}}=2.434$. Note that for the $h^{-1}_{11/2}$ level we have
used the position suggested in Refs. \cite{gene00a,mach02}.

The results for the odd-odd nuclei $^{132}$Sb, $^{130}$Sb and the odd ones $^{133}$Te, $^{131}$Sb are presented in subsections A and B, respectively. All calculations have been performed using the OXBASH shell-model code \cite{OXBASH}.

\subsection{The odd-odd nuclei $^{132}$Sb and $^{130}$Sb}

The most appropriate system to study the proton-neutron interaction is $^{132}$Sb,
with one proton valence particle and one neutron valence hole. Experimental information on this nucleus is provided by the studies of Refs. \cite{ston89,mach95,bhat01}.
Two $\beta$-decaying isomers with $J^{\pi}=4^+$ and $8^-$ are known, which originate from the $\pi g_{7/2}$ $\nu d^{-1}_{3/2}$ and 
$\pi g_{7/2}$ $\nu h^{-1}_{11/2}$ configurations, respectively.
While the relative energy of these two states has not been determined, there are indications
that the $8^-$ state is located about 200 keV above the $4^+$ ground state \cite{ston89}.  
As regards the positive-parity states, besides the other three members of the 
$\pi g_{7/2}$ $\nu d^{-1}_{3/2}$ multiplet, some states originating from the
$\pi g_{7/2}$ $\nu s^{-1}_{1/2}$, $\pi d_{5/2}$ $\nu d^{-1}_{3/2}$, and  $\pi g_{7/2}$ $\nu d^{-1}_{5/2}$ configurations have been identified. However, only  
three other negative-parity states have been observed and they have   
not received firm spin assignment. In a very recent study \cite{bhat01}, several new transitions feeding the $8^-$ isomer have been revealed.  While most of these new states have been interpreted as excitations of the $^{132}$Sn core,
the $9^-$, $10^+$, and $11^+$ states at 1.025, 2.799, and 3.199 MeV are attributed to the 
$\pi g_{7/2}$ $\nu h^{-1}_{11/2}$ and $\pi h_{11/2}$ $\nu h^{-1}_{11/2}$ configurations.
In this context, we have found it interesting to also include in our study the $^{130}$Sb nucleus, for which five low-lying states with $J^\pi=5^-,6^-,7^-,8^-$ and $9^-$ have been identified \cite{walte94,gene02} and interpreted as members of the 
$\pi g_{7/2}$ $\nu h^{-1}_{11/2}$ multiplet. 
The comparison between theory and experiment for $^{130}$Sb offers therefore the opportunity to gain more information on the proton-neutron interaction.   

Several calculated multiplets for $^{132}$Sb are reported in Fig. 3 and compared with the existing experimental data. Note that in Fig. 3(a) all energies are relative to the $4^+$ state while in Fig. 3(b) they are relative to the $8^-$ state (our calculations predict that the $8^-$ state lies 226 keV above the  $4^+$ ground state). It should also be noted that in the $\pi h_{11/2}$ $\nu h^{-1}_{11/2}$ multiplet we have not included the $0^+$, $1^+$, and $2^+$ states, since no positive-parity state with one of these angular momenta was found to be dominated by this component. For the $3^+$, $4^+$, and $5^+$ members of this multiplet, the percentage of configurations other than the dominant one ranges from 23\% to 33 \%. All the other states reported in Fig. 3 have a leading component whose percentage is at least 85\%.

As regards the comparison between theory and experiment, we see that the calculated energies are in good agreement with
the observed ones. In fact, the discrepancies are all in the order of few tens of keV, the only exceptions being the
$1^+$  and $9^-$ states of the $\pi d_{5/2}$ $\nu d^{-1}_{3/2}$ and $\pi g_{7/2}$ $\nu h^{-1}_{11/2}$ multiplets, which come about 300 keV above and 200 keV below their experimental counterparts, respectively.  

It is evident from Fig. 3 that a main feature of all these multiplets (obviously leaving aside  the doublet $\pi g_{7/2}$ $\nu s^{-1}_{1/2}$) is that the states with minimum and maximum $J$ have the highest excitation energy, while the state with next to the  highest $J$ is the
lowest one. This pattern is in agreement with the experimental one for the 
$\pi g_{7/2}$ $\nu d^{-1}_{3/2}$ multiplet and the few experimental data available for the other multiplets also go in the same direction.
Before moving to $^{130}$Sb, it is worth mentioning that in a very recent paper \cite{mach02} some preliminary results are reported concerning the states of the $\pi g_{7/2}$ $\nu h^{-1}_{11/2}$ multiplet. These results are consistent with the predictions of our calculations.

The calculated $\pi g_{7/2}$ $\nu h^{-1}_{11/2}$ multiplet for $^{130}$Sb is reported and compared  with the available experimental data in Fig. 4. For the experimental levels at 0.085, 0.112, and 0.145 MeV we adopt the $J^{\pi}=5^-$, $6^-$, and $7^-$ assignment proposed in Ref. \cite{walte94}, the $6^-$ state having been also observed in a more recent experiment
\cite{gene02}. For the level at 0.870 MeV we take  the spin-parity value $J^{\pi}=9^{-}$ \cite{gene02}.  We see that the calculated energies practically overlap the experimental ones, thus providing further support to our predictions for $^{132}$Sb.  As regards the wave functions of the states reported in Fig. 4, we find that they are indeed dominated by the $ph$ configuration  $\pi g_{7/2}$ $\nu h^{-1}_{11/2}$ while the remaining two 
neutron holes give rise to a zero-coupled pair occupying the $d_{3/2}$, $h_{11/2}$, and $s_{1/2}$ levels. 

Based on the above agreement between theory and experiment,  we have tried to clarify the nature of the low-lying positive-parity states of $^{130}$Sb, none of them having received a firm spin assignment . In Table I all the calculated and experimental levels
\cite{nndc} up to 0.350 MeV are reported while in the higher-energy region we only include the four levels observed in Ref. \cite{gene02}. 
The negative-parity states already presented in Fig. 4 are also included for the sake of completeness.
First, we note that our calculations predict for the ground state $J^\pi=4^+$ instead of $8^-$ as experimentally observed. However, in the observed spectrum the $(4,5)^+$ state is located  only 5 keV above the ground state.
As regards the four experimental levels at 0.005, 0.068, 0.075, and 0.267 MeV, we may identify them with our low-lying $4^+$, $3^+$, $5^+$, and $2^+$  states, which is supported by the behavior of the   $\pi g_{7/2}$ $\nu d^{-1}_{3/2}$ multiplet in $^{132}$Sb.
The level at 0.346 MeV can be associated with the calculated second 
$3^+$ state arising from the $\pi g_{7/2}$ $\nu s^{-1}_{1/2}$ configuration. For the three highest states reported in Table I,
we confirm the spin-parity assignment proposed in Ref. \cite{gene02} as well as the interpretation of the $11^+$ and $13^+$ states
as arising from the $\pi g_{7/2}$ $\nu d^{-1}_{3/2}h ^{-2}_{11/2} $ configuration. For the $10^-$ state, we predict an admixture of different three-neutron hole configurations with the spectator proton in the $g_{7/2}$ level.

\subsection{The odd nuclei $^{133}$Te and $^{131}$Sb}

Let us now come to the results for $^{133}$Te and $^{131}$Sb. Both these nuclei have been the subject of recent experimental studies \cite{bhat01,hwan02,gene00b}, where high-spin states of a particularly simple structure were identified.  The calculated spectra of $^{133}$Te and $^{131}$Sb are compared with the experimental ones in Figs. 5 and 6, where we also report the dominant configurations. We include  all the observed and calculated states up to 1.3 and 1.8 MeV for the former and latter nucleus, respectively. The experimental levels in this energy region are taken from Ref. \cite{nndc}, except the $J^{\pi}=\frac{19}{2}^{-}$ in $^{131}$Sb which is reported in Ref. \cite{gene00b}. In the higher energy regions, where several states with unknown or ambiguous spin and parity have been observed, 
we only include for both nuclei the high-spin states reported in \cite{bhat01,hwan02,gene00b}. 
As regards the calculated spectra, we report the states which can be safely associated with the experimental ones. In addition, we also show one (yrast) or two (yrast and yrare) states with $J\geq 13/2$ 
which can be unambiguously identified as belonging to the $\pi g_{7/2}^{2}$ $\nu h^{-1}_{11/2}$, $\pi g_{7/2} d_{5/2}$ $\nu h^{-1}_{11/2}$,  $\pi g_{7/2}$ $\nu h^{-2}_{11/2}$ configurations for $^{133}$Te and to the $\pi g_{7/2} $ $\nu h^{-2}_{11/2}$, $\pi g_{7/2} \nu d^{-1}_{3/2}$ $h^{-1}_{11/2}$ configurations for $^{131}$Sb.

For $^{133}$Te we find that the six lowest states shown in Fig. 5 originate from the  $\pi g_{7/2}^{2}$ $\nu d^{-1}_{3/2}$ configuration and have a 
percentage of the dominant component ranging from 66\% to 89\%. While the proton wave functions of the ground and first
excited states are mainly of seniority-0 nature, the other four states result essentially from the coupling of a neutron hole in the $d_{3/2}$ level to the first $2^+$ of $^{134}$Te. The percentage of the dominant component in the
other three groups of states reported in Fig. 5 is at least 90\%, the only exceptions being the $\frac{17}{2}^-$ at
2.4 MeV (53\%)  and the two $\frac{21}{2}^-$ states. As regards these two latter states, they contain almost the same admixture of the $\pi g_{7/2} d_{5/2}$ $\nu h ^{-1}_{11/2}$ and  $\pi g_{7/2}^{2}$ $\nu h ^{-1}_{11/2}$ configurations.
The three groups characterized by a neutron hole in the $h_{11/2}$ level correspond just to those existing in the experimental spectrum of $^{134}$Te, which arise from the $\pi g_{7/2}^{2}$, $\pi g_{7/2}d_{5/2}$ 
and $\pi g_{7/2}h_{11/2}$ configurations.  This reflects the effect of the proton-proton 
effective interaction, while the proton-neutron interaction is responsible for the arrangement of the states inside each group.
As regards the quantitative agreement between theory and experiment, we see that all the observed excitation energies are very well reproduced by our calculations, the discrepancies being well below 100 keV for most of the states.

For $^{131}$Sb, we find that none of the first seven positive-parity states is dominated by a single configuration. However, it turns out that the ground and first excited states result essentially from the coupling of the valence  proton in the $g_{7/2}$ and $d_{5/2}$  levels to  the ground state of $^{130}$Sn, while in the other five states the $g_{7/2}$ valence proton is coupled to the first $2^+$ state. The states of the next 
two groups shown in Fig. 6 are instead dominated by a single configuration. They correspond to the  
$\nu d_{3/2}^{-1} h_{11/2}^{-1} $ and $\nu h_{11/2}^{-2}$ multiplets in $^{130}$Sn, with the proton in the $g_{7/2}$ level. From Fig. 6 we see that up to 1.8 MeV a one-to-one correspondence can be established between the experimental and calculated spectra. The
experimental  state at 1.76 MeV with unknown spin and parity may be identified with our $\frac{9}{2}^-$ state at 1.80 MeV. It should be mentioned that some other states, which are not reported  in Fig. 6,  are predicted by the theory just above the  $\frac{9}{2}^-$ one.
As it was the case for $^{133}$Te, the quantitative agreement is quite satisfactory, the discrepancies  being larger than 100 keV only for the excitation energies
of the $\frac{3}{2}^+$ and $\frac{11}{2}^+$ states.

\section{Discussion and conclusions}

In this work, we have performed shell-model calculations for the four nuclei
$^{132}$Sb, $^{130}$Sb, $^{133}$Te and $^{131}$Sb, employing an effective interaction derived from the CD-Bonn nucleon-nucleon potential. This has been done within the framework of a new approach \cite{bogner02} to shell-model effective interactions which provides an advantageous alternative to the usual Brueckner $G$-matrix method. The effective interaction in the proton-neutron channel has been explicitly calculated in the $ph$ formalism.

We have shown that all the experimental data available for these nuclei are well reproduced by our calculations. These data, however, are still rather scanty and we have found it challenging to make predictions which may stimulate further experimental efforts to study these neutron-rich nuclei lying well away from the valley of stability. 

In this connection, of great interest are the proton-neutron hole multiplets. A relevant outcome of our calculations is that the highest- and lowest-spin members of each multiplet are well separated from the other states which lie very close in energy (see Figs. 3 and 4). This behavior is completely consistent with the experimental data (only one multiplet, however, is completely known) and is quite similar to that observed more than thirty years ago for the multiplets in the heavier particle-hole nucleus $^{208}$Bi \cite{kuo68,moin69}. Also, it is worth noting that in all of our calculated multiplets the state of spin ($j_\pi + j_\nu -1$) is the lowest, in agreement with the predictions of the Brennan-Bernstein coupling rule \cite{bren60}. 

To conclude this discussion, let us make some comments about our previous calculations on $^{132}$Sb \cite{andr99}, which were briefly outlined  in the Introduction. To start with, we should point out that they also led to a good agreement with the available experimental data. In Table II the excitation energies of the low-lying positive-parity states obtained from the previous and present calculations are reported and compared with the experimental ones. We see that the differences between the results of the two calculations do not exceed few tens of keV. It turns out, however, that significant differences exist for some states which have no experimental counterpart, in particular the $J=0^+$ states. Of course, these are to be ascribed to the differences in the two calculations, namely the
$NN$ potential we start from and the derivation of $V_{\rm eff}$ which implies
having two different closed cores, $^{100}$Sn and $^{132}$Sn. Since a main feature of the present calculations is the derivation of the proton-neutron effective interaction in the particle-hole formalism, we have found it interesting to calculate the $ph$ matrix elements starting from the $pp$ ones of Ref.  \cite{andr99} by means
of the well-known Pandya transformation \cite{pand56} and compare them with those obtained directly in the $ph$ formalism. To make this comparison free from ambiguites, the latter have been calculated starting from the Bonn-A $NN$ potential. By way of illustration, the matrix elements $<\pi g_{7/2} \nu d^{-1}_{3/2};J|V_{\rm eff}|\pi g_{7/2} \nu d^{-1}_{3/2}; J>$ and $<\pi h_{11/2} \nu h^{-1}_{11/2}; J|V_{\rm eff}|\pi h_{11/2} \nu h^{-1}_{11/2}; J>$ obtained in the two cases are reported in Table III. We note that the differences are within 150 keV in all cases, with the exception of the $J=0$ matrix elements which differ by about 700 keV. These findings are in line with those of the study \cite{west69}, where it was shown that the $pp$-$ph$ transformation of Pandya is an approximation to a more general many-particle relation.

\begin{acknowledgments}
This work was supported in part by the Italian Ministero dell'Universit\`a
e della Ricerca Scientifica e Tecnologica (MURST) and by the U.S. DOE 
Grant No.~DE-FG02-88ER40388.
\end{acknowledgments}

%\clearpage

\begin{figure}
\caption{First- and second-order one-body $\hat Q$-box diagrams.}
\end{figure}

\begin{figure}
\caption{First- and second-order two-body $\hat Q$-box diagrams.}
\end{figure}

\begin{figure}
\caption{Proton particle-neutron hole multiplets in $^{132}$Sb.
The theoretical results are represented by open circles while the experimental data by solid triangles. The lines are drawn to connect the points. See text for comments.}
\end{figure}

\begin{figure}
\caption{Same as Fig. 3, but for the $\pi g_{7/2}$ $\nu h^{-1}_{11/2}$ multiplet in  $^{130}$Sb.}
\end{figure}

\begin{figure}
\caption{Experimental and calculated spectra of $^{133}$Te.}
\end{figure}

\begin{figure}
\caption{Experimental and calculated spectra of $^{131}$Sb.}
\end{figure}

%\clearpage

\begin{table}
\caption{Experimental and calculated excitation energies (in MeV)  
for $^{130}$Sb.}

\begin{ruledtabular}
\begin{tabular}{clcl}
\multicolumn{2} {c} {Expt.} & \multicolumn{2} {c} {Calc.} \\
$J^{\pi}$ & $E$ & $J^{\pi}$ & $E$ \\
\hline
$8^-$&$0.0$ & $4^+$ & 0.0\\
$(4,5)^+$&$0.005$ & $3^+$ & 0.069\\
$(+)$&$0.068$ & $8^-$ & 0.092\\
$(3,4,5)^+$&$0.075$ & $5^+$ & 0.102\\
$6^-$&$0.085$ & $6^-$ & 0.155\\
$5^-$&$0.112$ & $5^-$ & 0.183\\
& & $3^-$ & 0.217\\
$7^{-}$&$0.145$ & $7^-$ & 0.235\\
$(2,3)^+$&$0.267$ &$2^+$ & 0.239 \\
$(2^{+},3^{+})$&$0.346$ & $4^-$ & 0.293\\
& & $3^+$ & 0.304\\
$(9^-)$&$0.870$ & $9^-$ & 0.831\\
$(10^-)$&$1.143$ & $10^-$ & 1.293\\
$(11^+)$&$1.508$ & $11^+$ & 1.593\\
$(13^+)$&$1.545$ & $13^+$ & 1.630\\

\end{tabular}
\end{ruledtabular}
\end{table}

\begin{table}
\caption{Excitation energies of low-lying positive-parity states in
$^{132}$Sb. See text for comments.}

\begin{ruledtabular}
\begin{tabular}{cccc}
$J^{\pi}$ & Expt. & Present work & Previous work\\
\hline
$4^+$ & 0.0   & 0.0 & 0.0 \\
$3^+$ & 0.086 & 0.089 & 0.090 \\
$5^+$ & 0.163 & 0.218 & 0.241 \\
$2^+$ & 0.426 & 0.418 & 0.452 \\
$3^+$ & 0.529 & 0.536 & 0.530 \\
$2^+$ & 1.078 & 1.168 & 1.089 \\
\end{tabular}
\end{ruledtabular}
\end{table}

\begin{table}
\caption{Diagonal $ph$ matrix elements of V$_{\rm eff}$ calculated (a) directly
in the $ph$ formalism and (b) from $pp$ matrix elements through the Pandya
transformation. In both cases use has been made of the Bonn-A $NN$ potential.
See text for comments.}

\begin{ruledtabular}
\begin{tabular}{cccc}
Configuration & $J$ & (a) & (b)  \\
\hline
$ \pi g_{7/2} \nu d^{-1}_{3/2}$ & 2 & 0.573 & 0.475 \\
 & 3 & 0.140 & 0.095 \\
 & 4 & 0.017 & -0.051 \\
 & 5 & 0.249 & 0.188 \\
$ \pi h_{11/2} \nu h^{-1}_{11/2}$ & 0& 3.345 & 2.660 \\
 & 1 & 1.752 & 1.904 \\
 & 2 & 0.818 & 0.775 \\
 & 3 & 0.467 & 0.507 \\
 & 4 & 0.352 & 0.293 \\
 & 5 & 0.232 & 0.205 \\
 & 6 & 0.177 & 0.132 \\
 & 7 & 0.130 & 0.096 \\
 & 8 & 0.071 & 0.037 \\
 & 9 & 0.139 & 0.108 \\
 & 10 & -0.013 & -0.025 \\
 & 11 & 0.497 & 0.522 \\
\end{tabular}
\end{ruledtabular}
\end{table}

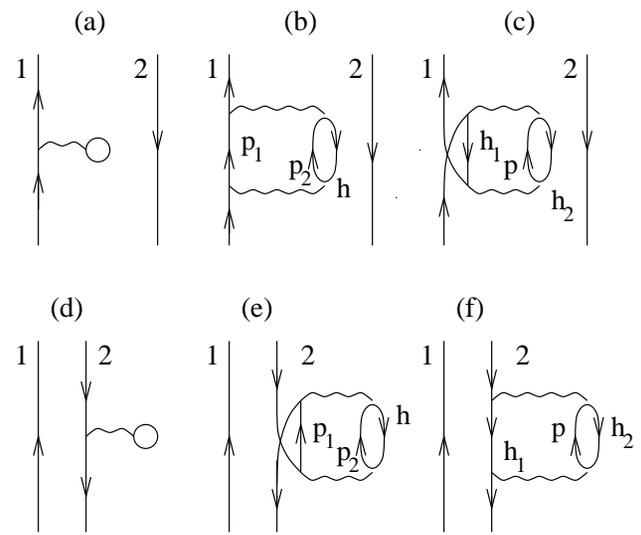
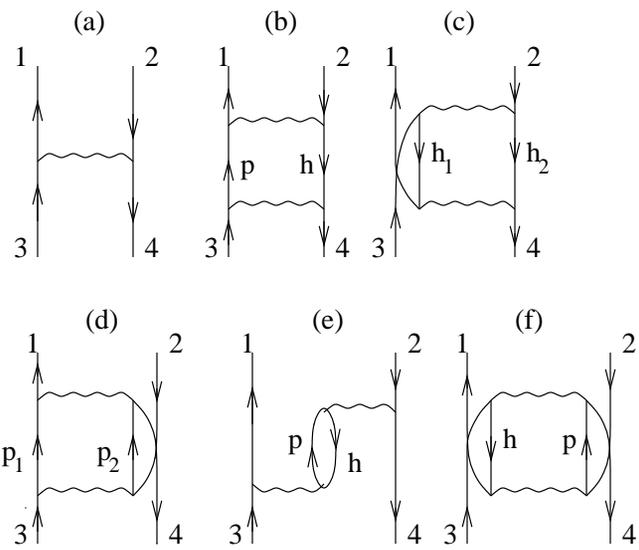
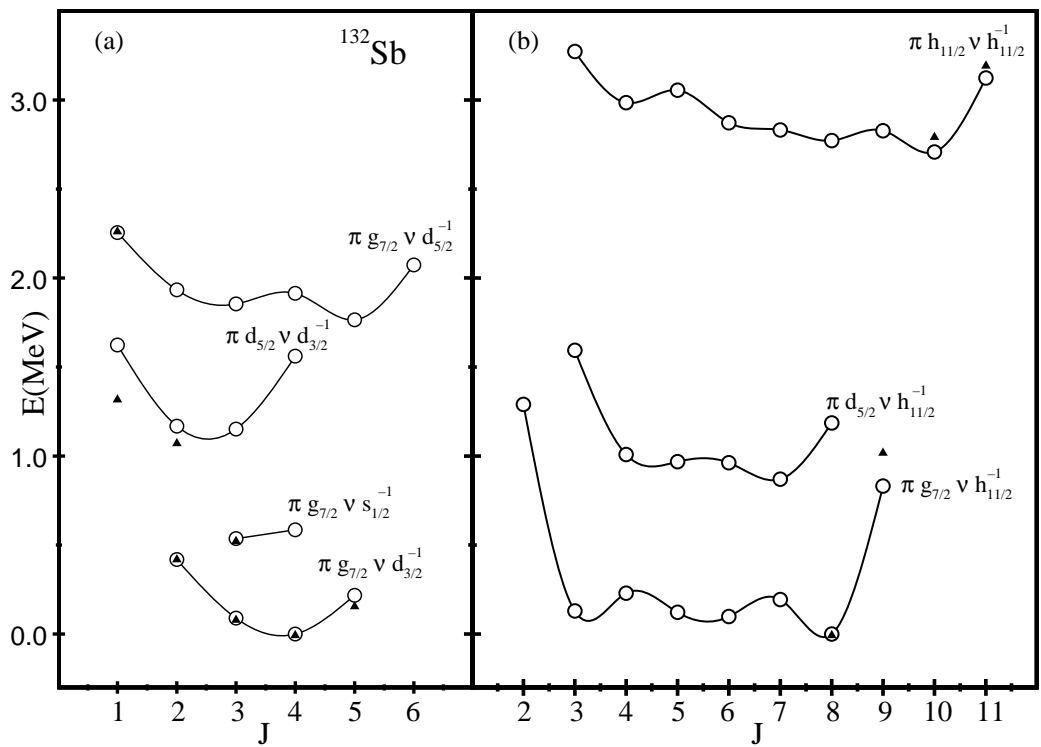
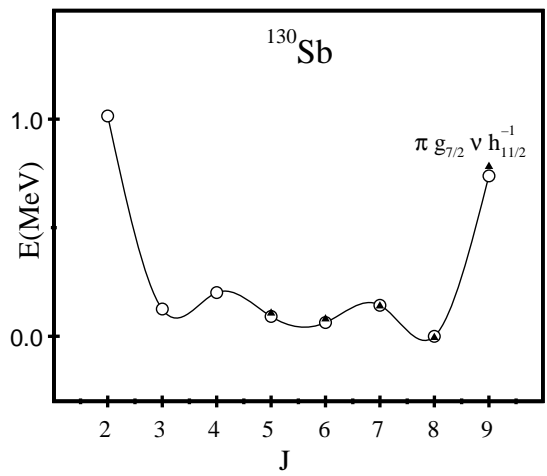
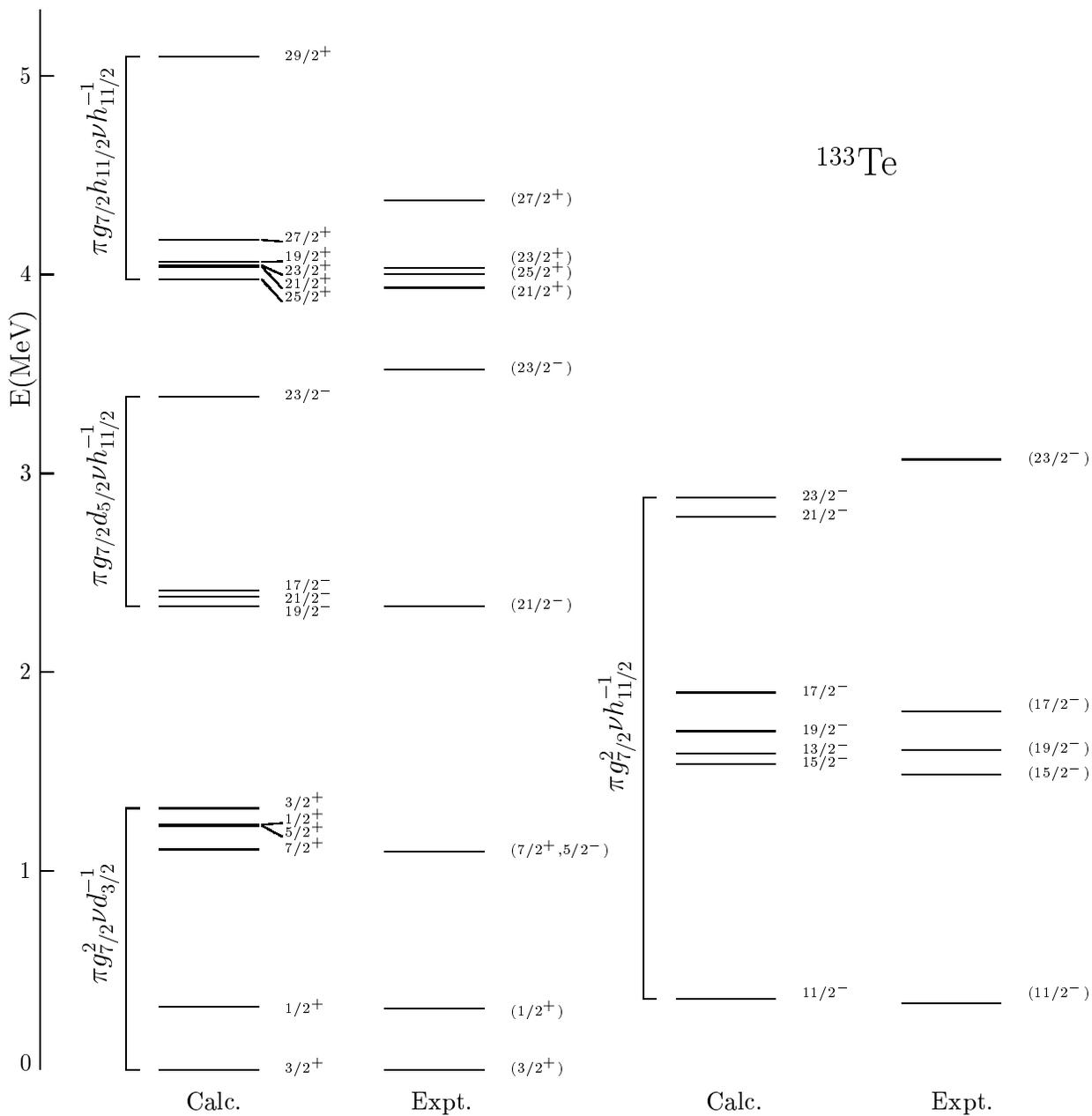
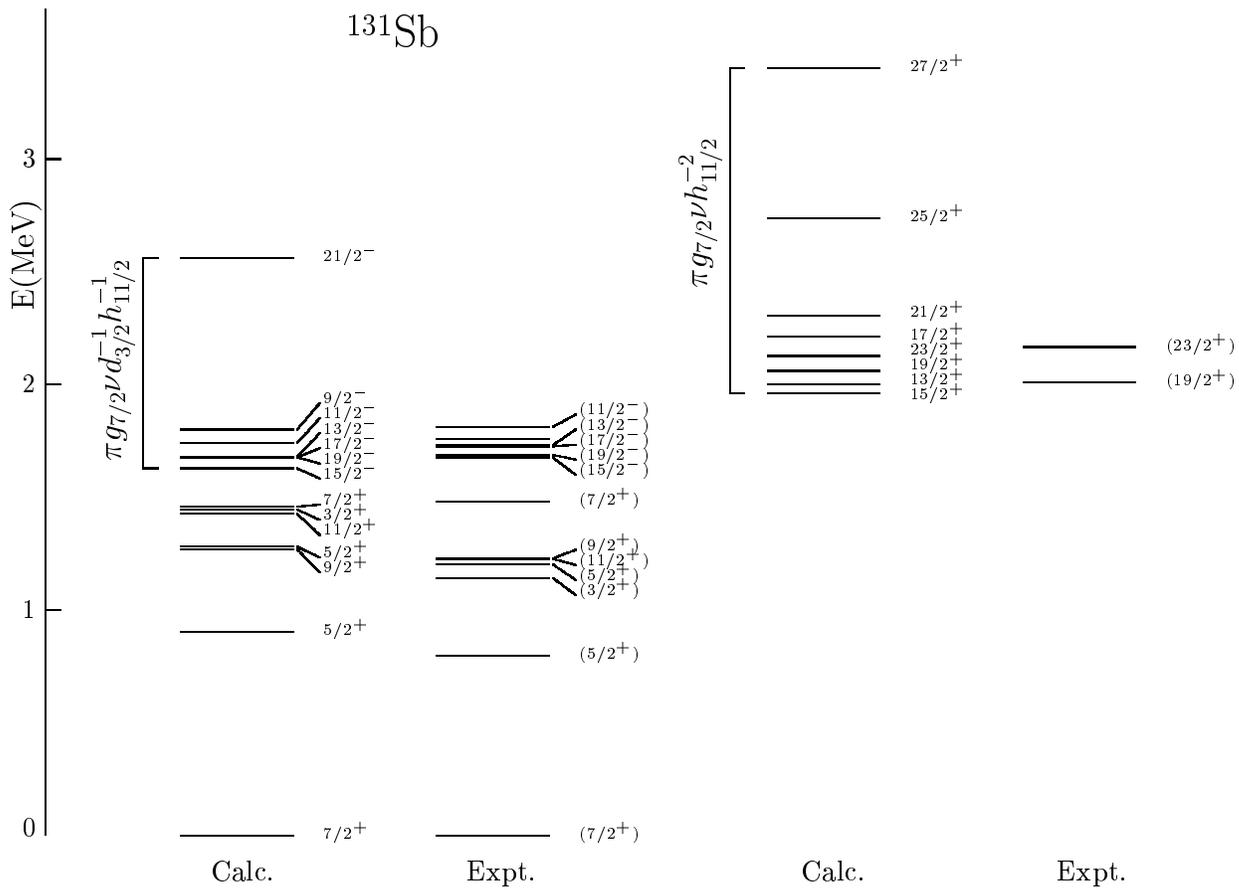
\end{document}